\documentclass{aastex}\usepackage{emulateapj5}
\received{}
\accepted{}
\journalid{}{}
\articleid{}{}
\shortauthors{Wiltshire}
\shorttitle{}

\begin{document}
%\parskip 3pt
%----------------------------------------------------------------
% Local definitions
\def\w#1{\,\hbox{#1}} \def\Deriv#1#2#3{{#1#3\over#1#2}}
\def\etal{{\it et al}.} \def\e{{\rm e}} \def\de{\delta}
\def\dd{{\rm d}} \def\ds{\dd s} \def\ep{\epsilon} \def\de{\delta}
\def\th{\theta} \def\ph{\phi} \def\et{\eta}
\def\ts{t} \def\tc{\tau} \def\rh{\rho}
\def\goesas{\mathop{\sim}\limits} \def\OM{\widetilde\Omega} \def\gam{\gamma}
\def\gc{\gam\Z0} \def\bH{\bar H_i} \def\be{\beta} \def\zb{{\bar z}}
\def\Z#1{_{\lower2pt\hbox{$\scriptstyle#1$}}} \def\ac{\atil\Z0} \def\al{\alpha}
\def\X#1{_{\lower2pt\hbox{$\scriptscriptstyle#1$}}} \def\tn{\tc\Z0}
\def\Om{\Omega_m} \def\aphys{a} \def\atil{\widetilde a}
\def\Oc{\OM\Z0} \def\omi{{\OM}_i} \def\Hec{H\ws{eff$\;$0}}
\def\Hb{\widetilde H} \def\MM#1{{\cal M}^{#1}} \def\Vm{V(\varphi_0)}
\def\bz{{\bar z}} \def\bb{\beta'} \def\wc{w\Z0} \def\Hm{H\Z0}
\font\sevenrm=cmr7 \def\ws#1{_{\hbox{\sevenrm #1}}}
\def\Ws#1{\Z{\,\hbox{\sevenrm #1}}}
\def\PRL#1{Phys.\ Rev.\ Lett.\ {\bf#1}} \def\PR#1{Phys.\ Rev.\ {\bf#1}}
\def\ApJ#1{Astrophys.\ J.\ {\bf#1}} \def\AsJ#1{Astron.\ J.\ {\bf#1}}
\def\frn#1#2{{\textstyle{#1\over#2}}} \def\SS{${\cal S}$}
\def\lsim{\mathop{\hbox{${\lower3.8pt\hbox{$<$}}\atop{\raise0.2pt\hbox{$\sim$}}
$}}} \def\gsim{\mathop{\hbox{${\lower3.8pt\hbox{$>$}}\atop{\raise0.2pt\hbox{$
\sim$}}$}}} \def\dL{d\Z L} \def\dA{d\Z A}
\def\Hef{H\ws{eff}} \def\kmsMpc{\w{km}\;\w{sec}^{-1}\w{Mpc}^{-1}}
\def\FRW{(\ref{FRWopen1}), (\ref{FRWopen2})}
\def\frw{(\ref{FRWopen1})--(\ref{FRWopen4})}
\def\KMNR{Kolb, Matarrese, Notari and Riotto}
\def\beq{\begin{equation}} \def\eeq{\end{equation}}
\def\bea{\begin{eqnarray}} \def\eea{\end{eqnarray}}
%----------------------------------------------------------------
\title{Viable inhomogeneous model universe without dark energy from primordial
inflation}

\author{David~L.~Wiltshire}%\altaffilmark{1}}}

%\altaffiltext{1}
\affil{Department of Physics \& Astronomy,
University of Canterbury, Private Bag 4800, Christchurch, New Zealand}

\begin{abstract}
A new model of the observed universe, using solutions to the full Einstein
equations, is developed from the hypothesis that our observable
universe is an underdense bubble, with an internally inhomogeneous fractal
bubble distribution of bound matter systems, in a spatially flat bulk universe.
It is argued on the basis of primordial inflation and resulting structure
formation, that the clocks of the isotropic observers in average galaxies
coincide with clocks defined by the true surfaces of matter homogeneity of
the bulk universe, rather than the comoving clocks
at average spatial positions in the underdense bubble geometry, which are in
voids. This understanding requires a systematic reanalysis of all observed
quantities in cosmology. I begin such a reanalysis by giving a model of the
average geometry of the universe, which depends on two measured parameters:
the present matter density parameter, $\Om$, and the Hubble constant, $\Hm$.
The observable universe is not accelerating. Nonetheless,
inferred luminosity distances are larger than na\"{\i}vely expected,
in accord with the evidence of distant type Ia supernovae.
The predicted age of the universe is $15.3\pm0.7$ Gyr.
The expansion age is larger than in competing models, and may account
for observed structure formation at large redshifts.
\end{abstract}
%\pacs{98.80.-k, 98.80.Es, 98.80.Cq, 98.80.Bp }
\maketitle
\section{Introduction}
Observations in the past decade have been interpreted as suggesting
that 70\% of the matter--energy density in the universe at the present
epoch is in the form of a smooth vacuum energy, or ``dark energy'', which
does not clump gravitationally. This is supported by two powerful
independent lines of observation. Firstly, type Ia supernovae in distant
galaxies (Perlmutter \etal\ 1998, 1999; Riess \etal\ 1998, 2004) are dimmer
than would be expected in standard Friedmann--Robertson--Walker
(FRW) models, especially when it is noted that many independent dynamical
estimates of the present clumped mass fraction, $\Om$, suggest values
of order 20--30\%. Secondly, observations of the power spectrum of
primordial anisotropies in the cosmic microwave background radiation (CMBR),
most recently by the WMAP satellite (Bennett \etal\ 2003), indicate that the
universe appears to be spatially flat on the largest of scales.

A cosmological constant, or alternatively dynamical dark energy, is most
commonly invoked to explain the observed cosmological parameters, even though
a fundamental origin for such dark energy remains one of the profoundest
mysteries of modern physics. However, even in the presence of dark
energy at the present epoch, a number of problems remain. One of the
most significant problems is that the epoch of reionization measured
by WMAP appears at a redshift of order $z\goesas20^{+10}_{-9}$,
indicating that the first stars formed much earlier than conventional models
of structure formation would suggest. The detection of complex galaxies at
relatively large redshifts compounds the conundrum (see, e.g., Cimatti
\etal\ 2004, Glazebrook \etal\ 2004).

In recent work, \KMNR\ (2005) have proposed a profoundly different
resolution of the ``cosmological constant problem''. They reason that
primordial inflation will have produced density perturbations many times
larger than the present horizon volume. We should not view the observable
universe as typical of the universe on scales larger than our particle
horizon, and the values of cosmological parameters
observed should not necessarily be taken as typical of the
whole. Furthermore, they suggest that our present observations might be
compatible with the observed universe being an underdense bubble in
an otherwise spatially flat $k=0$ FRW universe with an energy density
$\Omega\Ws{TOT}=1$ in ordinary matter. Such assumptions are indeed
supported by detailed numerical and analytic calculations made in
inflationary models a decade ago (Linde, Linde and Mezhlumian 1996).

The possibility of inhomogeneously defined clock rates is commonly
accepted in studies of inhomogeneous cosmologies, such as the
Lema\^{\i}tre--Tolman--Bondi (LTB) models. (For a review
see Krasi\'nski (1997).) In this {\it Letter}, I will argue, however,
that primordial inflation can give rise to a particular structure
in inhomogeneous cosmologies, which allows for a homogeneous cosmic time
on the scales of the first bound systems which form, which differs from
the ``comoving'' time parameter of the average late epoch geometry. This
novel feature is what distinguishes the present model from previous studies
of inhomogeneous cosmologies, as it leads to a new solution of the
{\it fitting problem} (Ellis and Stoeger 1987).

\section{Observers, clocks and the fitting problem in inflationary cosmology}
It is a consequence of the inflationary paradigm that a spectrum of initially
small density perturbations is stretched to all observable scales within
our past light cone, and also to scales beyond our particle horizon. The fact
that this is true for the past light cone is well supported by the CMBR. The
hypothesis that such perturbations should extend to super--horizon scales
is a feature of most inflationary models, independent of their details.

One important realisation is the fact that since primordial inflation
ended at a finite very early time, the scale of super--horizon sized modes,
although huge, must have a cut--off at an upper bound. Beyond that scale
we assume the universe is described by a spatially flat bulk metric
\beq\label{FRWflat}
\ds^2\ws{bulk} = - \dd\tc^2 + \bar a^2(\tc)(\dd x^2+\dd y^2+\dd z^2),
\eeq
where $\bar a(\tc)=\bar a_i (\tc/\tc_i)^{2/[3(1+w)]}$, and we use units
with $c=1$. The precise bulk equation of state $P\ws{bulk}=w\rho\ws{bulk}$
proves to be inconsequential if $w>-1$, though in accord with the principles
of the model we assume only ordinary matter and radiation.

Although the observable universe
will undoubtedly be embedded in many regions of under- and over-density,
like the smallest figure inside a Russian doll, it is nonetheless reasonable
to assume that provided the density perturbation immediately containing our
observed universe extends sufficiently beyond our horizon then there is a
super-horizon sized underdense bubble containing the observable universe, with
matter density {\it equal to the average matter density we measure},
which we can model as a super--horizon sized underdense region, \SS.

The matter distribution inside \SS\ is assumed to be {\it inhomogeneous and
``fractal''}, in accord with calculations from primordial inflation
(Linde, Linde and Mezhlumian 1996)
and in accord with all the evidence of observations of the actual universe,
with its large-scale structure of streaming motions
of clusters of galaxies, bubbles and voids.
Nonetheless, despite this inhomogeneity we do observe an average isotropic
Hubble flow. Thus at some level the first step in solving the fitting
problem (Ellis and Stoeger 1987) must involve the approximation of the
inhomogeneous
geometry of \SS\ by an average spacetime which depends on a single cosmic
scale factor, viz., the spatially open FRW geometry
\beq\label{FRWopen1}
\dd\widetilde s^2 = - \dd\ts^2 + \atil^2(\ts)\left[{\dd r^2\over1+r^2}
+r^2(\dd\th^2+\sin^2\th\,\dd\ph^2)\right],
\eeq
where parametrically in terms of conformal time, $\et$,
\bea
\atil&=&{a_i\omi\over2(1-\omi)}(\cosh\et-1)\,,\nonumber\\
H_{i}\ts&=&{\omi\over2(1-\omi)^{3/2}}(\sinh\et-\et)\,,
\label{FRWopen2}\eea
where $\omi$ is an initial density parameter of the bubble, \SS,
and $a_i$ a constant to be set.

Usually the first step after writing down the metric (\ref{FRWopen1}) is
to identify comoving observers in this geometry with idealised isotropic
observers, namely {\it observers who measure no dipole anisotropy in the
CMBR}. One identifies typical stars in typical galaxies with such observers,
apart from the small effects of peculiar velocities. However,
even in an homogeneous isotropic FRW model this implicitly assumes
a solution of the fitting problem in terms of a complicated matching of
asymptotic scales to relate the clocks on geodesics in
bound systems, where space is not expanding, to the scale of comoving
observers in the Hubble flow, where space is expanding. It is assumed
without question that such clocks can be identified.

In general relativity with the particular self--similar inhomogeneous
geometry that arises from the evolution of the %scale-invariant spectrum of
density perturbations of %that one would find in an underdense bubble after
primordial inflation, the standard solution of the fitting problem is not
appropriate, as I argue in detail in a subsequent paper (Wiltshire 2005),
henceforth Ref.~I. In brief, while the underdense bubble, \SS, is on its
largest scale a single perturbation away from the average bulk density,
{\it due to the scale-invariance of the inflationary spectrum, perturbations
on the smaller and intermediate scales that give rise to the first structures
nonetheless have the same statistical distribution as the bulk universe, with
a mean density distributed about that of the bulk}. The particular structure
of inhomogeneity we see today, with its large voids, results in part from
a {\it``particle horizon volume selection bias''} in the initial density
perturbations (see Ref.~I).

While the metric (\ref{FRWopen1}) is still valid for describing the clocks of
an observer at an average {\it spatial} position, in an
inhomogeneous underdense bubble such {\it average spatial positions are in
voids and do not coincide with stars and galaxies}. On the other hand,
the first bound systems of stars and star clusters which aggregate to
typical galaxies, break away from the Hubble flow at early epochs
when the average local density of matter in their past light cones is close
to the average density of the bulk. Such
systems retain fragments of the bulk hypersurface geometry. A clock in a
system which has had an approximate stationary Killing vector since the
epoch of break away, thus measures the bulk cosmic time parameter, $\tc$,
which is ``frozen in'' from an earlier epoch in the evolution of the universe.
Initial perturbations with the selection--biased
underdensity of the bubble, \SS, evolve to form voids.

Given that structure forms in a bottom--up manner, a {\it``temporal''}
Copernican principle operates within \SS\ in the sense that {\it as average
isotropic observers we assume we are located in a bound system
formed from those density perturbations which broke away from the Hubble flow
when the geometry within their past light cone was indistinguishable from that
of the bulk surfaces of homogeneity}. Rather than occupying an average
position on a spatial hypersurface, we find ourselves in a bound system that
formed from matter which broke from an almost homogeneous Hubble flow at an
average epoch. At the present epoch, these same average galaxies, are located
in clusters in bubble walls surrounding local voids, in a self--similar
hierarchical structure, which we loosely call {\it fractal}.

Since recent expansion of space occurs primarily in voids, the question of
who does or does not measure an isotropic CMBR depends not only on local
peculiar velocities but also on whether an observer's
line of sight to the surface of last scattering in any direction on the sky
averages over the same number and volume distribution of voids and
bubble walls, once such structures form. For observers in
average galaxies in bubble walls or observers within small voids one would
expect a roughly isotropic distribution of bubbles and voids within the
self--similar hierarchical matter distribution, and hence an almost
isotropic CMBR. Clock rates are determined by local geometry rather than
by isotropy, or otherwise, of the CMBR at any location.

To define the local cosmic time, $\tc$, in bound systems in average galaxies,
as we have done, in reference to bulk hypersurfaces of matter homogeneity
which stretch to regions beyond our present particle horizon may seem puzzling
at first. However, prior to inflation, such regions {\it were in causal
contact with the observed universe}. Thus the same basic processes of
inflation that lead to isotropy of the CMBR also lead, via structure
formation from scale--invariant perturbations, to a definition of inertial
frames of observers in average galaxies, namely a new variant of Mach's
principle. (See Ref.~I for further discussion.)

Since general relativity is a local theory, the average geometry in our
first--order solution of the fitting problem is described
by \FRW, but insofar as measurements, including our own, are referred to
the frame of average galaxies we must refer them to a coordinates
related to those of (\ref{FRWopen1}) by a non--trivial
lapse function $\gam(\tc)\equiv\Deriv{\dd}\tc{\ts}$, so that
(\ref{FRWopen1}) becomes
\beq\label{FRWopen3}
\dd\widetilde s^2 = \gam^2(\tc)\ds^2\,,
\eeq
where
\beq\label{FRWopen4}
\ds^2= -\dd\tc^2 +
\aphys^2(\tc)\left[{\dd r^2\over1+r^2}+r^2(\dd\th^2+\sin^2\th\,\dd\ph^2)
\right],\eeq
and $\aphys\equiv \gam^{-1}\atil$. {\it Since quantities such as the
luminosity distance involve null geodesics, for many calculations we can use
the conformally related geometry (\ref{FRWopen4}).} It is not necessary to
solve for the geodesics of observers, defined by local
geometry. Only null geodesics probe
the vast cosmological scales over which the average geometry (\ref{FRWopen1})
is more relevant than other scales in the actual inhomogeneous geometry.
Spacelike intervals are never directly measured.
Thus by fiat, once we insist on writing the average cosmological geometry
in a ``synchronous gauge'' adapted to the proper time, $\tc$, of the local
clocks of observers in average galaxies,
then the geometry (\ref{FRWopen4}) also comes to define our rods
when specifying areas, volumes and densities on the largest scales.
The quantity $\gam(\tc)$ leads to a gravitational time dilation between
average galaxies in the gravitational wells of the bubble walls and
the ideal comoving clocks in empty voids, as further discussed in
Ref I.

To analyse some quantities will require a specification of the
entire inhomogeneous fractal geometry within \SS. This may be possible
in the context of a LTB model. However, given the hierarchical nature of the
inhomogeneity the LTB mass function would be much more complicated than
those of the simple single- or few-void LTB models that have typically been
studied to date. Initially, we are only interested in properties defined by
the average Hubble flow, although we might expect some modifications of
cosmological parameters due to the different range of scales and
expansion rates among local
voids (Tomita 2001). While solving for intermediate scales, perhaps in terms
of a LTB model, will ultimately be necessary, for the purpose of the first
approximation in the fitting problem, it suffices to construct a
``spherical expansion model'' of the underdense
region \SS, in parallel to the well--known
``spherical collapse model'' (Kolb \& Turner 1990).

\section{Spherical expansion model}
Following the standard approach, we assume the initial density parameter
is set sufficiently early that it is very close to unity:
$\omi=1-\de_i$, $0<\de_i\ll 1$.
Furthermore, since the universe has $\omi\simeq1$ initially we require
the comoving scales to match at that epoch. This means we set $\bar a_i
\simeq a_i$, $\gam_i\simeq1$, $\ts_i\simeq\tc_i$ and $H_i\simeq2/(3\tc_i)$.

There are various possible Hubble
parameters to take into account:\hfil\break %\begin{itemize}\item
{\bf(i)} the background Hubble parameter $\bar H=2/[3(1+w)\tc]$ of the
metric (\ref{FRWflat}) which cannot be directly measured in \SS;\hfil\break
{\bf(ii)} the Hubble parameter of comoving observers in (\ref{FRWopen1}),
measurable only by unbound observers in average voids
\beq
\Hb(\ts)\equiv{1\over\atil}\Deriv{\dd}{\ts}\atil\quad ;
\label{Hvoid}\eeq
%\item
%\item
{\bf(iii)} the physical Hubble parameter we measure
\beq
H(\tc)\equiv{1\over\aphys}\Deriv{\dd}{\tc}\aphys\quad;
\label{Hphys}\eeq
{\bf(iv)} an effective (unmeasured) parameter
\beq
\Hef(\tc)\equiv{1\over\atil}\Deriv{\dd}{\tc}\atil\quad,
\label{Heff}\eeq
which is useful for the purposes of calculations.

By our assumptions, while all these parameters coincide at early times,
as the universe expands they may differ.
As observers in average galaxies, using null geodesics we measure
the cosmological geometry (\ref{FRWopen4})
referred to global cosmic time,
$\tc$, not to the parameter $\ts$. We simply need to determine $\ts(\tc)$
to determine cosmological quantities.

Using Eqs.\ \frw, the bulk universe scale factor, $\bar a=\bar a_i\left(\frn1n
\bar H_i\tc\right)^n$, where $n=2/[3(1+w)]$ and $\bar H_i=3nH_i/2$, the
density parameter is defined as
\beq
\OM=\omi\left(a_i\over\atil\right)^3\left(\bar a\over\bar a_i\right)^{2/n}=
{18{H_i}^2(1-\omi)^3\tc^2\over{\omi}^2(\cosh\et-1)^3}\,.
\label{bom}\eeq
Comparing this with the standard parametric expression $\OM(\et)=2
(\cosh\et-1)/\sinh^2\et$ derived from \FRW, we obtain a parametric
relation for $\tc(\et)$,
\beq
H_i\tc={\omi(\cosh\et-1)^2\over3(1-\omi)^{3/2}\sinh\et}\,.
\label{age1}\eeq
The effective parameter (\ref{Heff}) is found to be
\beq
\Hef(\et)%&=&{3H_i(1-\omi)^{3/2}\sinh^3\et\over\omi^{3/2}(\cosh\et-1)^2
%(\sinh^2\et+\cosh\et-1)}\,\nonumber\\
={\Hec(1-\Oc)^{3/2}(\Oc+2)(\cosh\et+1)^{3/2}
\over\Oc(\cosh\et-1)^{3/2}(\cosh\et+2)}
\label{hubble_eff}\eeq
where $\Oc=2/(1+\cosh\et\Z0)$, and a subscript zero refers to
the present epoch. The physically measured Hubble parameter
(\ref{Hphys}) is
\bea
H(\et)={(\cosh^2\et+2\cosh\et+3)\Hef(\et)
\over(\cosh\et+1)(\cosh\et+2)}\,.
\label{hubble}\eea
and the measured Hubble constant, $\Hm$, is related to $\Hec$ by
\beq\Hm=\left(2+{\Oc}^2\over 2+\Oc\right)\Hec\,.\eeq
Observe that at early times, $\et\goesas0$, $\Hef\goesas\Hb$ and
$H\goesas\Hb$, as expected but
at late times, $H\goesas\Hef\goesas\frn32\Hb$.
The lapse function is given by
\beq
\gam(\et)=\Deriv{\dd}\tc{\ts}={\Hef\over\Hb}={3(\cosh\et+1)\over
2(\cosh\et+2)}\,. \label{gamma}\eeq
At the present epoch $\gc=3/(2+\Oc)$.

We must also be careful to note that the locally measured density parameter,
differs from (\ref{bom}) by a volume factor according to
$\Omega(\tc)=\gam^3(\tc)\OM(\tc)$, so that the present epoch measured
matter density fraction is
\beq
\Om={27\Oc\over(2+\Oc)^3}\,,
\eeq
with inverse $\Oc=6\Om^{-1/2}\sin\left[\frac\pi{6}-\frac{1}{3}\cos^{-1}
(\Om^{1/2})\right]-2$.

The expansion age (\ref{age1}) is usefully rewritten as
\beq
\tc(\et)={\Oc(2+\Oc^2)(\cosh\et-1)^{3/2}\over\Hm(1-\Oc)^{3/2}(\Oc+2)^2
(\cosh\et+1)^{1/2}}\,.
\label{age2}\eeq

The physically measured deceleration parameter $q(\tc)=-H^{-2}\ddot a/a=
-1-\dot H/H^2$, is given by
\beq
q(\et)={7\cosh^2\et+10\cosh\et+1\over(\cosh^2\et+2\cosh\et+3)^2}\,.
\label{decel}\eeq
It is equal to the bulk deceleration parameter $\bar q=\frn12$ at early
times $\et=0$, but at late times as $\et\to\infty$, $q\to0$,
so that it is small but positive at the present epoch.
This is just as expected from a model without dark energy, and
contradicts the claims of cosmic acceleration by Kolb \etal\ (2005).
However, we must recall that the supernovae measurements involve luminosity
{\it distances}, and the {\it interpretation} of cosmic acceleration depends
on the time parameter assumed in taking derivatives. The present model
may be recognised as mimicking a Milne universe at nearby redshifts, and
this still provides a reasonable fit to present SneIa data
(Carter \etal\ 2005).

\section{Observational tests}
To compare with observations Eqs.\ (\ref{hubble})--(\ref{decel}) need to be
expressed in terms of the observed cosmological redshift, $z$.
Great care is needed at this point in identifying physical quantities.
It proves simplest to always use the line element (\ref{FRWopen3}),
(\ref{FRWopen4}) remembering that $\tc$ corresponds to clock time for
observers in average galaxies. It follows that
\beq
1+z={a\Z0\over a}={\ac\gam\over\atil\gam\Z0}={(1-\Oc)(2+\Oc)
(\cosh\et+1)\over\Oc(\cosh\et+2)(\cosh\et-1)}\,.
\label{redshift}\eeq
The physical solution to eqn.~(\ref{redshift}) is
\bea
\cosh\et&=&{-1\over2}+{(1-\Oc)(2+\Oc)\over2\Oc(z+1)}\nonumber\\
&+&{\sqrt{\Oc z[9\Oc z -2{\Oc}^2+16\Oc+4]
+({\Oc}^2+2)^2}\over2\Oc(z+1)}\,.\nonumber\\
\label{solutn}
\eea

We are now ready for our first cosmological tests.
The luminosity distance is readily computed in the standard fashion and
is found to be $\dL=(1+z){\gc}^{-1}\ac\sinh(\et-\et\Z0)$, and since
$\sqrt{1-\Oc}=\gc/(\ac\Hec)$ it follows that
\beq
\Hm\dL %=\left(2+{\Oc}^2\over2+\Oc\right)\Hec\dL
={(1+z)(2+{\Oc}^2)\over\Oc(2+\Oc)}\left\{2\cosh\et -{(2-\Oc)\over\sqrt{1-\Oc}}
\sinh\et\right\}\,.
\label{dL}\eeq
As shown by Carter \etal\ (2005) the model gives a good fit to the
supernova data, with or without non--baryonic cold dark matter in addition
to baryonic matter.

The expansion age is given by substituting (\ref{solutn}) in (\ref{age2}). It
gives a present day age of the universe of
\beq\tn={2(2+\Oc^2)\over(2+\Oc)^2\Hm}\,.\eeq
%$\tn=2(2+\Oc^2)/[(2+\Oc)^2\Hm]$.
Using the best--fit value (Carter \etal\ 2005) for $\Hm=62.7^{+1.1}_{-1.7}
\kmsMpc$, the age of the universe is $\tn=15.0\pm0.4$
Gyr for $\Om=0.25\pm0.5$, or alternatively $\tn=15.4^{+0.6}_{-0.4}$ Gyr
for a universe with only baryonic matter,
$\Om=0.075\pm0.055$, using the recalibrated primordial nucleosynthesis
bounds of Ref.~I.

Importantly, the expansion age is significantly larger at large $z$: with
$\Hm=62.7^{+1.1}_{-1.7}\kmsMpc$, for $\Om=0.25\pm0.5$ we find $\tc=4.3\pm0.3$
Gyr at $z=2$, $\tc=1.45^{+0.15}_{-0.12}$ Gyr at $z=6$, and $\tc=0.31\pm0.04$
Gyr at $z=20$. For $\Om=0.075\pm0.055$ then $\tc=4.9^{+0.2}_{-0.3}$ Gyr at
$z=2$, $\tc=1.94^{+0.26}_{-0.21}$ Gyr at $z=6$, and $\tc=0.51^{+0.17}_{-0.10}$
Gyr at $z=20$. This would buy precious time
for structure formation early on, and help to explain the redshift
of the reionization epoch (Bennett \etal\ 2003), and observations of
``early'' formation of structure (Cimatti \etal\ 2004,
Glazebrook \etal\ 2004).

\section{Conclusion}
As a model cosmology, the present {\it Fractal Bubble Universe} is remarkable
in that its gross features depend only on two already observed parameters,
and the results obtained thus far are reasonably promising. If correct, then
all observed quantities in cosmology must be systematically reanalysed. The
determination of some quantities will require further development of our
understanding of the inhomogeneous matter distribution within \SS, including
the largest scale local voids (Tomita 2001), in particular. However,
the present model offers a clear framework for calculations, as is
demonstrated in Ref.~I, where further steps
are taken in the reanalysis of the hot Big Bang and observational
implications for the CMBR.

It would be ironic that Einstein's idea concerning the
irrelevance of the cosmological constant, and his idea of the importance
of Mach's principle, may prove to both be right in understanding
the universe.

\medskip\noindent
{\bf Acknowledgement} This work was supported by the
Mardsen Fund of the Royal Society of New Zealand. I thank Jorma Louko,
Jenni Adams, Roy Kerr and other members of the UC gravity group for
helpful comments.

%--------- References ---------%

%-------------------------------------------------------------------

\begin{thebibliography}{99}

\bibitem[Bennett \etal(2003)]{wmap}
Bennett, C.L., \etal\ 2003,
%{\it``First Year Wilkinson Microwave Anisotropy Probe (WMAP) Observations:
%Preliminary Maps and Basic Results''},
\apjs\ {\bf148}, 1. %[arXiv:astro-ph/0302207]
%%CITATION = ASTRO-PH 0302207;%%

\bibitem[Carter \etal(2005)]{paper1}
Carter, B.M.N., Leith, B.M., Ng, S.C.C., Nielsen, A.B. and Wiltshire, D.L.
2005,
%{\it``Exact model universe fits type IA supernovae data with no cosmic
%acceleration''},
astro-ph/0504192.

\bibitem[Cimatti \etal(2004)]{Cimatti}
Cimatti, A., \etal\ 2004,
%{\it``Old galaxies in the young universe''},
\nat\ {\bf430}, 184.
%%CITATION = NATUA,430,184;%%

\bibitem[Ellis \& Stoeger(1987)]{ES}
Ellis, G.F.R. and Stoeger, W. 1987,
%{\it``The fitting problem in cosmology''},
Class.\ Quantum Grav.\ {\bf4}, 1697.
%%CITATION = CQGRD,4,1697;%%

\bibitem[Glazebrook \etal(2004)]{Glazebrook}
Glazebrook, K., \etal\ 2004,
%{\it``A high abundance of massive galaxies 3.6 billion years after the
%Big Bang''},
\nat\ {\bf430}, 181.
%%CITATION = NATUA,430,181;%%

\bibitem[Kolb(2005)]{Kolb}
Kolb, E.W., Matarrese, S., Notari, A. and Riotto, A. 2005,
%{\it``Primordial inflation explains why the universe is accelerating
%today''},
arXiv:hep-th/0503117.
%%CITATION = HEP-TH 0503117;%%

\bibitem[Kolb \& Turner(1990)]{KT}
Kolb, E.W. and M.~S.~Turner, M.S. 1990,
{\it``The Early Universe''}, (Addison--Wesley, Reading, Mass.)

\bibitem[Krasi\'nski(1997)]{Krasinski}
Krasi\'nski, A. 1997,
{\it``Inhomogeneous Cosmological Models''}, (Cambridge Univ. Press).

\bibitem[Linde, Linde \& Mezhlumian(1996)]{LLM}
Linde, A.D., Linde, D., and Mezhlumian, A. 1996,
%{\it``Nonperturbative amplifications of inhomogeneities in a self-reproducing
%universe''},
\prd\ {\bf 54}, 2504. %[arXiv:gr-qc/9601005].
%%CITATION = GR-QC 9601005;%%

\bibitem[Perlmutter \etal(1997)]{Perl1}
Perlmutter, S., \etal\ 1997,
%{\it``Measurements of the cosmological parameters $\Omega$ and $\Lambda$
%from the first seven supernovae at $z\ge 0.35$''},
\apj\ {\bf483}, 565; %[arXiv:astro-ph/9608192]
%%CITATION = ASJOA,483,565;%%

\bibitem[Perlmutter \etal(1999)]{Perl2}
Perlmutter S., \etal\ 1999,
%{\it``Measurements of $\Omega$ and $\Lambda$ from 42 high-redshift
%supernovae''},
\apj\ {\bf 517}, 565. %[arXiv:astro-ph/9812133]
%%CITATION = ASJOA,517,565;%%

\bibitem[Riess \etal(1998)]{Riess1}
Riess, A.G., \etal\ 1998,
%{\it``Observational evidence from supernovae for an accelerating universe and
%a cosmological constant''},
\aj\ {\bf116}, 1009; %[arXiv:astro-ph/9805201]
%%CITATION = ANJOA,116,1009;%%

\bibitem[Riess \etal(2004)]{Riess2}
Riess, A.G., \etal\ 2004, %[Supernova Search Team Collaboration],
%{\it``Type Ia supernova discoveries at z$>$1 from the Hubble
%Space Telescope: Evidence for past deceleration and constraints
%on dark energy evolution''},
\apj\ {\bf 607}, 665. %[arXiv:astro-ph/0402512]
%%CITATION = ASTRO-PH 0402512;%%

\bibitem[Tomita]{Tomita}
Tomita, K. 2001,
%{\it``A local void and the accelerating universe''},
\mnras\ {\bf326}, 287. %[arXiv:astro-ph/0011484].
%%CITATION = ASTRO-PH 0011484;%%

\bibitem[Wiltshire(2005)]{paper2}
Wiltshire, D.L. 2005,
in preparation.
\end{thebibliography}
\end{document}